# Behavior Pattern and Compiled Information Based Performance Prediction in MOOCs

Shaojie Qu
Network and Information Center
*Beijing Institute of Technology*
Beijing, China
qushaojie@bit.edu.cn

Kan Li (corresponding author)
School of Computer Science
*Beijing Institute of Technology*
Beijing, China
likan@bit.edu.cn

Zheyi Fan
School of Mathematics and Statistics
*Beijing Institute of Technology*
Beijing, China
1120162096@bit.edu.cn

Sisi Wu
School of Mathematics and Statistics
*Beijing Institute of Technology*
Beijing, China
wuweisbit@126.com

Xinyi Liu
School of Mathematics and Statistics
*Beijing Institute of Technology*
Beijing, China
Xyliu@126.com

Zhiguo Huang
School of Mathematics and Statistics
*Beijing Institute of Technology*
Beijing, China
1120162565@bit.edu.cn

*Abstract*—**With the development of MOOCs (massive open online courses), increasingly more subjects can be studied online. Researchers currently show growing interest in the field of MOOCs, including dropout prediction, cheating detection and achievement prediction. Previous studies on achievement prediction mainly focused on** students' video and forum behaviors, and few researchers have considered how well students perform their assignments. In this paper, we choose a C programming course as the experimental subject, which involved 1528 students. This paper mainly focuses on the students' accomplishment behaviors in programming assignments and compiled information from programming assignments. In this paper, feature sequences are extracted from the logs according to submission times, submission order and plagiarism. The experimental results show that the students who did not pass the exam had obvious sequence patterns but that the students who passed the test did not have an obvious sequence pattern. Then, we extract 23 features from the compiled information of students' programming assignments and select the most distinguishing features to predict the students' performances. The experimental results show that we can obtain an accuracy rate of 70.49% for predicting students' performances.

## I. INTRODUCTION

MOOCs have achieved great success in lifelong learning and students' learning. Now, researchers are increasingly studying the field of MOOCs, including dropout prediction[1], cheating detection[2] and achievement prediction[3]. Previous research on achievement prediction has mainly focused on demographic characteristics, grades, course participation and mood[4], while our research is based on students' overall behaviors patterns when handing in assignments. We study more about long-term, unconscious behavior pattern of certain student.

In this paper, we choose a C programming course as the experimental subject. We focus on the behavior of when students submit their programming assignments and then compile information about their programming assignments. Students' behavior of submitting assignments may reflect their learning motivation and attitude, while the compiled information of students' programming assignments may reflect their learning status. We hope to find the behavior patterns of students with different performance levels and to predict students' performances through the compiled information of students' assignments.

We adopted a method of sequence patterns to find the behavior pattern of students. Corresponding feature sequences were extracted from the logs according to submission times, submission order and plagiarism. The experimental results showed that the students who did not pass the final exam had obvious sequence patterns, but the students who passed the final exam did not have obvious sequence patterns. Compiled information about students' programming assignments was extracted, and distinguishing features were selected to predict the students' performances. The experimental results showed that we could obtain an accuracy rate of 70.49% for predicting students who could pass the exam.

The arrangement of this paper is as follows: the second section introduces the relevant work, the third section addresses the problems we need to study, and the fourth section introduces the data and describes the methods of the results. In fifth section, the experimental results are analyzed. Section six summarizes the conclusion and proposes future work

## II. RELATED WORKS

### A. MOOCs

The vast amounts of logs stored in MOOCs makes studying student learning patterns possible. The usual research approach is to use students' participation in MOOCs to predict their final scores. According to the early performance of the students, Qiujie Li[5] divided students into four subgroups, including rounders, listeners, auditors and disengagers. This study mainly focused on finding the correlation between behavior engagement and cognitive engagement with grades. Research has shown that achievement can be predicted through early participation. However, the relationship between early participation and achievement varies by subgroups.

Gökhan Akçapınar[6] used text information to determine whether plagiarism behavior occurred and further explored automated feedback through text mining, which can significantly decrease online plagiarism behavior.

In 2014, Paulo Blikstein[7] used methods from machine learning to find a pattern in the data of 154,000 code snapshots from 370 students and to predict their final exam grades. The results showed that the process-based metric was more predictive for final exams than the midterm grades. In 2014, Petri Ihantola[8] studied the correlation between students' behaviors and the perceived difficulty of students' programming tasks by using behavioral data in JAVA courses. The analysis showed that both the time spent on the task and the number of programming events were moderately to highly correlate with perceived difficulty.

## B. Sequential pattern mining

Sequential pattern mining has developed rapidly and has been applied in various fields[9]. In 1995, sequential pattern mining (SPM) was first proposed by Agrawal et al. to discover frequent sequences. AprioriAll, AprioriSome, and Dynamicsome[10] were defined in this paper. The generalized sequential pattern mining (GSP)[11] algorithm was proposed in the following year and worked better than the AprioriAll algorithm in generating candidate sequences. However, these algorithms require scanning the sequence database repeatedly to find all frequent sequences, and they are not efficient in dealing with long sequence patterns.

Zaki in 2001 proposed the SPADE[12] algorithm for higher efficiency, which divided the sequence into three equivalence classes and decomposed the original problem. As a result, we need to consider only three database scans, which increases mining efficiency. Ayres et al. introduced SPAM[13], an algorithm that first represented the original sequence database in vertical bitmap form, and then applied a novel depth-first search strategy to determine all sequence patterns. SPAM has been demonstrated to be far better than SPADE for large datasets. However, this algorithm has high memory requirements because it puts all sequences into main memory.

Pei et al.[14][15] constructed a projected database to avoid scanning the database repeatedly and proposed algorithms based on the previous ones called FreeSpan and PrefixSpan. FreeSpan and PrefixSpan outperform Apriori-based algorithms because they work only on the projected database without multiple scans of the original database. However, these algorithms need to create projected databases, which consumes large amounts of time and memory space.

More methods for sequential patterns, such as closed sequential pattern mining[16], and multidimensional sequential pattern mining[17], have been developed in recent years.

## III. RESEARCH QUESTIONS

*Question A:*

What are the behavior patterns associated with assignments for students with different performance levels?

*Question B:*

How can we predict students' performances with the compiled information from programming assignments?

## IV. DATA AND METHOD

### A. Data

We collected the logs of a C programming MOOC course, which involved 1528 students from non-information schools. We extracted features from programming assignments and compiled information.

We divided the 69 programming assignments into 14 groups according to weekly order and content. For each student, we constructed three feature sequences based on the submission order, submission times and plagiarism times. The length of each of these sequences was 14, corresponding to 14 groups of assignments. The value of each element in the submission order sequence was the average submission order of all assignments in the corresponding group. The element value of the submission times sequence was the average submission time of assignments in the corresponding group, and the element value of the plagiarism sequence was the sum of the plagiarism times in the corresponding group.

We extracted 23 features from compiled information, including the number of compiled errors for 22 types and the number of compiled successes.

### B. Method for question A

We divided the students into two groups according to whether they passed or failed in the final exam; then, we devised a GSP-based method to find the behavior pattern of each group.

Assuming the range of element values in the sequence is {A,B,C,D,E}, if the sequence ADBCE can match 70% of all the sequences, then the prefix ADBC must match at least 70% of all the sequences. Then, when we search forward for sequence patterns, if a sequence cannot match 70% of the sequence, then the extended sequence on the basis of the sequence must not be a pattern sequence that meets the condition. We first exhaust all possible sequences with a length of 1, delete the sequences that do not meet the requirement of 70% matching degree, and then judge whether the accuracy of the screened sequences in predicting student performance is greater than 70%. We retain only the sequences with both an accuracy rate and a recall rate greater than 70%. To run the algorithm, we need to standardize the data.

We use the submission times sequence to reflect students' abilities. We assume that in group j of assignments $X_j$ is the average submission times for all students and $X_{ij}$ is the submission times for one student. Then we use the following formula to standardize the value in the sequence.

$$DR_{ij} \text{ (Difference rate)} = (X_{ij} - X_j) / X_{ij} \quad (1)$$

If $DR_{ij} < -0.5$, the corresponding element value is -2. If $-0.5 < DR_{ij} < 0$, the corresponding element value is -1. If $0 < DR_{ij} < 0.5$, the corresponding element value is 1. If $0.5 < DR_{ij}$, the corresponding element value is 2. If $DR_{ij} = 0$, the corresponding element value is 0.

We use the submission order sequence to reflect students' learning attitudes and initiatives. When the students' average submission order is between 0 and 500, the standardized value is 1; when the average submission order is between 501 and 1000, the standardized value is 2; when the average submission order is greater than 1001, the standardized value is 3.

We use the plagiarism sequence to reflect students' effort. When plagiarism times are 0, the standardized value is 0; when plagiarism times are between 1 and 2, the standardized value is 1; when plagiarism times are more than 2, the standardized value is 2.

After the completion of sequence standardization, we use algorithms to obtain the sequence patterns for students with different performances.

*C. Method for question B*

Ten of these features extracted from compiled information are shown in TABLE I with their meanings:

TABLE I. FEATURES AND MEANINGS

| Features | Meaning |
|---|---|
| None | No errors at all |
| Syntax error | Illegal statement in code |
| Redefinition of main | Main function repeatedly defined |
| Undeclared | The variable is not declared |
| Invalid value | Wrong data type or data size |
| Stray | Additional symbols appears |
| Invalid operands | Invalid operands to binary |
| Not a function | No correlation function defined |
| Conflicting | Inconsistent declaration of function |
| Not use struct | Invalid use of 'struct data' |

After extracting these 23 features, we use the random forest feature selection method to rank the features by importance. Then, we use multilayer perception (MLP) with three hidden layers to predict students' performance. We trained with 80% of the data and predicted the remaining 20%. Students are considered to be failed if their scores are less than 60. According to the ranking of feature importance, we add features to the optimal feature set from high importance to low importance and train them with MLP. When the accuracy is no longer improved, we stop adding features to obtain the optimal feature set.

V. RESULTS

*A. Results for question A*

We use the GSP-based algorithm on three sequences for two groups of students separately. The experimental results show that the students who failed the final exam have some similar behavior patterns, but the students who passed the final exam do not match these behavior patterns. This result means that all roads lead to Rome but that failure is always the same.

For submission times, we collected the following behavior patterns in TABLE II. (*) means that one or more elements can be added to the pattern in the sequence. For example, sequence '-2,2,1,1,2,-2,-1,-1,-2,1,-2,1,2,-1' follows sequential pattern (*)2(*)2(*)-2(*)-2(*), as there are two '-2' followed by two '2' in this sequence while (*) means one or more arbitrary data.

TABLE II. PATTERNS FOR SUBMISSION TIMES

| Sequential Pattern | Accuracy | Recall Rate |
|---|---|---|
| (*) 2 (*) 2 (*) -2 (*) -2 (*) | 70.01% | 81.44% |
| (*) -2 (*) -2 (*) -2 (*) -2 (*) | 71.19% | 75.78% |
| (*) 2 (*) 2 (*) 2 (*) 2 (*) -2 (*) | 71.38% | 76.22% |

For the first sequence, the early submission times are significantly higher than the average number, indicating that students have made more attempts to complete the assignment. In the later assignments, the submission times are significantly lower than the average number of submissions. The later assignments were more difficult but required fewer submissions, meaning that the students had cheated and did not try to complete the assignments. In the second sequence, the submission times of students in at least 4 groups are far less than the average submission times of all students, indicating that students cheated in many groups of assignments. The third sequence reflects that students submitted more times than average in at least four groups. This result may mean that although students work hard, they have difficulty completing the course. A crossover phenomenon occurs among the students covered by the three sequences. Some students may possibly satisfy the three sequence patterns at the same time. At present, we cannot combine the three sequence patterns.

For the submission order sequence, we obtained the following two sequences for failed students in TABLE III.

TABLE III. PATTERNS FOR SUBMISSION ORDER

| Sequential Pattern | Accuracy | Recall Rate |
|---|---|---|
| (*) 3 (*) 3 (*) | 71.20% | 77.44% |
| (*) 3 (*) 3 (*) 3 (*) | 73.83% | 70.22% |

A value of 3 indicates that the student submitted his or her assignments after 1000 students. The sequence (*) 3 (*) 3 (*) means that for two groups of assignments, students submitted their assignments after 1000 students on average, which shows that the students are not positive enough to finish their assignments. The result of the submission order sequence indicates that the later students submit their program, the higher probability that the students would fall.

As shown in Fig. 1, the submission orders of the failing students are mostly 2 or 3, while the submission orders of the students with grade more than 90 are 1 or 2, which means that the excellent students have a more positive attitude towards learning, and the failing students submit their assignments later and have a more negative attitude towards learning.

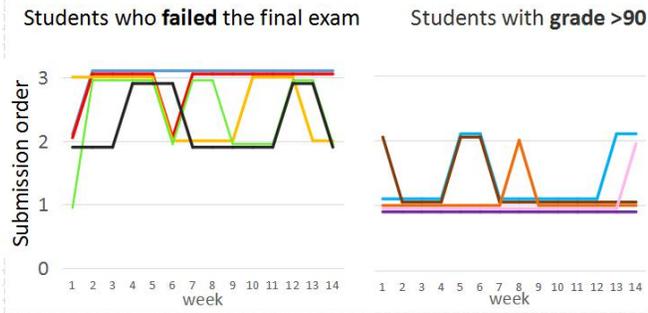

Fig. 1. Relationship between submission orders and final grade

In the sequence of plagiarism, we obtained two sequences for failed students in TABLE IV. If in one or two groups of assignments, a student commits serious plagiarism, then the probability is high that he or she will fail in the final exam.

TABLE IV. PATTERN FOR PLAGIARISM

| Sequential Pattern | Accuracy | Recall Rate |
|---|---|---|
| (*) 2 (*) | 73.49% | 85.00% |
| (*) 2 (*) 2 (*) | 78.73% | 70.22% |

*B. Results for question B*

We use the random forest method to evaluate the discrimination degree of all features. The experimental results are shown in TABLE V. The feature 'None' achieved the highest degree of differentiation, and students who failed the exam had fewer attempts rated 'None', meaning that students completed their programming assignments with fewer attempts. Because programming assignments have many complex test cases, passing a few tests means that students cope with many assignments. Syntax error indicates a basic syntax error, and those who often make syntax errors are failing at the basic knowledge level, which means that they are more likely to fail in the final exam.

TABLE V. FEATURE IMPORTANCE

| Features | Importance | Features | Importance |
|---|---|---|---|
| None | 0.473 | Undeclared | 0.068 |
| Syntax error | 0.198 | Invalid value | 0.038 |
| Redefinition of main | 0.076 | Stray | 0.023 |

We tested the selected features with some algorithms, and the experiment showed that the MLP algorithm achieved the highest accuracy; the accuracy of MLP in predicting whether students would pass the test reached 70.49%.

TABLE VI. COMPARATIVE RESULTS

| Method | Accuracy | Recall Rate |
|---|---|---|
| Naïve Bayes | 60.33% | 60.16% |
| Logic Regression | 62.30% | 49.14% |
| Support Vector Machine | 58.03% | 54.96% |
| MLP | 70.49% | 70.16% |

Experimental results show that we can select the most representative features from the compiled information and predict the performance of students.

## VI. CONCLUSION AND FUTURE WORK

In this paper, we take a C programming MOOC as the experimental object, and our goal is to predict the final performance of the students according to the assignments. First, we divide all assignments into 14 groups, and we obtain three separate sequences based on submitting times, submitting order and plagiarism. Through the analysis of the students' sequence patterns, we find that the students who passed the examination did not have an obvious sequence pattern, but students who failed the exam had some of the same behavior patterns. Second, we analyze the compiled information of students from programming assignments; we select discriminative features with a feature selection algorithm. Experiments show that students with different performances have obvious differences in specific compiled information. We use MLP to predict the performance of students and obtain an accuracy of 70.49%. The accuracy is not improved temporarily, as this is a work in progress. We will consider more features and introduce other algorithms to improve accuracy.

Although we find some behavior patterns in the assignment for the students who failed in the exam, the accuracy and recall rate of the behavior patterns have the potential to be further improved, and we will explore more effective behavior patterns in the future. At the same time, we have not combined these behavior patterns and the compiled information in our predicting framework. In the next step, we will combine the students' behavior patterns with the features extracted from the compiled information and make further predictions of the students' performance to achieve better prediction results.